# A 3.9 km baseline intensity interferometry photon counting experiment

Giampiero Naletto[*a,b], Luca Zampieri[c], Cesare Barbieri[c,d], Mauro Barbieri[e], Enrico Verroi[f], Gabriele Umbriaco[d], Paolo Favazza[d], Luigi Lessio[c], Giancarlo Farisato[c]

[a] Department of Information Engineering, Univ. of Padova, Via Gradenigo 6/B, 35131 Padova Italy
[b] CNR/IFN/LUXOR, Via Trasea 7, 35131 Padova, Italy
[c] INAF-Astronomical Observatory of Padova, Vicolo dell'Osservatorio 5, 35122 Padova, Italy
[d] Dept. of Physics and Astronomy, University of Padova, Vicolo Osservatorio 3, 35122 Padova, Italy
[e] Department of Physics, University of Atacama, Copayapu 485, Copiapo, Chile
[f] Inst. for Fundamental Physics and Applications (TIFPA), Via Sommarive 14, 38123 Povo TN, Italy

**ABSTRACT**

In the last years we have operated two very similar ultrafast photon counting photometers (Iqueye and Aqueye+) on different telescopes. The absolute time accuracy in time tagging the detected photon with these instruments is of the order of 500 ps for hours of observation, allowing us to obtain, for example, the most accurate ever light curve in visible light of the optical pulsars. Recently we adapted the two photometers for working together on two telescopes at Asiago (Italy), for realizing an Hanbury-Brown and Twiss Intensity Interferometry like experiment with two 3.9 km distant telescopes. In this paper we report about the status of the activity and on the very preliminary results of our first attempt to measure the photon intensity correlation.

**Keywords:** Quantum Astronomy, High Time Resolution Astrophysics, Intensity Interferometry, Photon counters

## 1. INTRODUCTION

The large majority of present astronomical instrumentation gets information about remote objects by the analysis of some properties of electromagnetic radiation. For example, by the measurement of the photon arrival directions through imaging systems, information can be retrieved about relative motion of stellar objects; from stellar intensity variation as measured by photometric systems, it is possible to infer the transit of exo-planets; from the energy of the incoming photons obtained with spectroscopic systems, it is possible to understand the type of the light source, and so on. However, there is other information somehow "hidden" in the light stream arriving from a stellar object. For example, individual photons may carry various amounts of orbital angular momentum in addition to their "regular" angular momentum associated with circular polarization [1][2]. Moreover, laboratory and theoretical studies in quantum optics [3] have demonstrated that both individual photons and groups of photons may carry additional information encoded in the (normalized) correlation functions of the electric field or the quantum field describing the photon gas [4],[5],[6], such as the physics of light emission (e.g. stimulated emission, as in a laser) or of propagation (e.g. whether photons have arrived directly from the source, or have undergone scatterings on their way to the detector) [7][8].

In principle, also light coming from celestial sources contains such an information: on this respect, we proposed some year ago to develop a *quantum* field in astronomy, with the aim of measuring these not yet fully exploited properties of light [9]. The instrument we proposed was named QuantEYE 0[11][12], the "quantum eye", to be applied to the largest telescope under study at the time, the ESO 100 m OverWhelmingly Large (OWL) telescope. In fact, one of the need of quantum astronomy is to have very large collecting areas, since what has to be measured is a high order effect, and a huge amount of photons is necessary to have a significant signal to noise ratio.

For realizing quantum astronomy, it is requested to measure the statistical properties of the time of arrival of photons coming from the same target with the best possible timing accuracy. Ideally, the time resolution should get as close as possible to the Heisenberg's uncertainty principle, that corresponds to ~1 ps for a spectral bandwidth of the order of 1 nm.

---

[*] giampiero.naletto@unipd.it, phone +39 049 8277646; http://www.dei.unipd.it/~naletto

Unfortunately, reaching such a stringent time resolution and accuracy is not possible with presently available photon counting detectors: in fact, with the fastest commercial single photon avalanche diodes (SPADs), at most a few tens of ps can be obtained as minimum time jitter on the photon detection [13]. Arrays of SPADs are now entering the market, but they are not a feasible solution to solve this issue since they just allow to increase the total count rate, but not actually the detector time resolution [14]. In any case, working in the nanosecond or sub-nanosecond regime still allows to measure significant second order effects, so the proposed QuantEYE was designed to work in the sub-nanosecond regime at GHz photon counting rates. Now OWL concept has been superseded by the more affordable 39 m European Extremely Large telescope (E-ELT), but our study still maintains its validity also if applied to E-ELT.

In order to test the proposed concept, we decided to acquire some experience by building two prototype instruments: Aqueye+ (Asiago Quantum Eye+)[15][16], for the Asiago 1.8 m Copernicus telescope, and Iqueye (Italian Quantum Eye) [17][18], for the ESO 3.5 m New Technology telescope (NTT) in La Silla (Chile). Both these instruments have demonstrated their extremely high time resolution performance, making the presently best time analysis of optical pulsars [19][20][21][22]. Concerning the possibility of measuring second order correlation in stream of photons coming from stellar objects, we already did some precursor experiments, in order to validate the technique, working on telescope sub-apertures [23][24]. Since these measurements have been done working with relatively small telescopes, we simply wanted to check the instrumentation, and to obtain not contradictory results with the expectations. What we tested was the capability to perform intensity interferometry (II) measurements working in photon counting regime in the optical range with an essentially null baseline. It is well known that II allows to obtain the angular size of the emitting source: this was demonstrated 40 years ago by Hanbury Brown and Twiss (HBT) with their intensity interferometer experiment [25][26], so far the only astronomical application dedicated to the measurement of second-order coherence of light. With their experiment, HBT actually exploited the *wave* nature of light, measuring the cross-correlation of the intensity fluctuations (second order spatial correlation of the electric field) of the star signal detected by two photomultipliers at the foci of two 6.5 m telescopes separated by a baseline up to a 180 meters. However, a stellar II experiment which exploits the *particle* nature of the light has not been performed yet. As shown in our QuantEYE study, now technology allows to realize such a measurement: if applied to a pair of large telescopes, or better to an array of large telescopes so working with several baselines, it is possible to obtain the source image reconstruction with unprecedented spatial resolution, down to microarcseconds with baselines of the order of km's [27].

In this paper we describe the possibility of testing for the first time a km-long baseline II experiment by detecting and very accurately time tagging (absolute time) visible photons. This is still a validation test, because of the limitation on both the relatively small telescope areas and the maximum count rate possible with the present electronics. However, the preliminary results obtained so far are very encouraging, and pave the way for a possible future application on a couple of large telescopes with a more efficient control electronics.

## 2. INTENSITY INTERFEROMETRY

Understanding II may not be trivial, because quantum optics textbooks are not always very clear, because each author has his own notation, because the phenomenon can be explained either in terms of classical waves, or light coherence, or particles properties, and so on. We are summarizing here the most relevant theoretical aspects of this phenomenon making reference to just a few classical papers on this topic, hoping to be able to provide a somehow ordered explanation; then we will conclude with some practical formulas for the measurement of this effect.

Let's start with a pair of classical definitions. Following [28], the cross correlation of two oscillating electrical fields at space-time points A($r_A$,$t_A$) and B($r_B$,$t_B$) is

$$\Gamma(r_A, t_A; r_B, t_B) = \langle E^*(r_A, t_A) E(r_B, t_B) \rangle \qquad (1)$$

where the $\langle ... \rangle$ operator denotes the "ensamble" average, which corresponds to the time average in the case of stationary and ergodic fields. Under these conditions, the time cross correlation depends only on the time difference $\tau = t_B - t_A$, and equation (1) becomes

$$\Gamma(r_A, r_B, \tau) = \langle E^*(r_A, t) E(r_B, t + \tau) \rangle \qquad (2)$$

also known as *mutual coherence function*. By suitable normalization, the *complex degree of coherence* of the light at the two points A and B is obtained as

$$\gamma(r_A, r_B, \tau) = \frac{\langle E^*(r_A, t) E(r_B, t+\tau)\rangle}{\sqrt{\langle |E^*(r_A,t)|^2\rangle \langle |E(r_B,t)|^2\rangle}} = \frac{\Gamma(r_A, r_B, \tau)}{\sqrt{\langle I(r_A,t) I(r_B,t)\rangle}} \tag{3}$$

An intensity interferometer measures the correlation between the *fluctuations* of intensities at two separated points in a partially coherent field. Following the discussion provided by [29], it can be found that the correlation between the intensities is given by

$$\langle I(r_A,t) I(r_B, t+\tau)\rangle = \langle I(r_A,t)\rangle \langle I(r_B,t)\rangle + \Gamma^2(r_A, r_B, \tau) = \langle I(r_A,t)\rangle \langle I(r_B,t)\rangle [1 + |\gamma(r_A, r_B, \tau)|^2]. \tag{4}$$

Making explicit the intensity fluctuations, $I(r_i, t) = \langle I(r_i, t)\rangle + \Delta I(r_i, t)$, it is easy to show that

$$\langle I(r_A,t) I(r_B, t+\tau)\rangle = \langle I(r_A,t)\rangle \langle I(r_B,t)\rangle + \langle \Delta I(r_A,t) \Delta I(r_B, t+\tau)\rangle. \tag{5}$$

From these two equations, denoting the spatial dependence with subscripts for a simpler notation, we obtain that the correlation between the fluctuations of intensities is simply

$$\langle \Delta I_A(t) \Delta I_B(t+\tau)\rangle = \langle I_A(t)\rangle \langle I_B(t)\rangle |\gamma_{AB}(\tau)|^2. \tag{6}$$

Finally, since this analysis refers to linearly polarized light while we are dealing with unpolarized light, a ½ correcting factor has to be introduced, providing that

$$\langle \Delta I_A(t) \Delta I_B(t+\tau)\rangle = \tfrac{1}{2}\langle I_A(t)\rangle \langle I_B(t)\rangle |\gamma_{AB}(\tau)|^2. \tag{7}$$

**2.1 The Hanbury Brown and Twiss approach**

In order to understand how this intensity fluctuation correlation can be used for the measurement of stellar diameters, we can take the simplified classical approach described in [26], which refers to the HBT II experiment in the case of polarized light. Let us consider two point sources $P_1$ and $P_2$ on the surface of a star, emitting white light independently one of the other, and let us collect some of this light by two separate optical detectors, each one having in front a narrow band spectral filter (same passband for the two). We can represent the wavefront of this light by means of the Fourier analysis as a large number of sinusoidal components with random amplitude and phase. Consider two of these Fourier components of the light, $E_1 \sin(\omega_1 t + \phi_1)$ from $P_1$ and $E_2 \sin(\omega_2 t + \phi_2)$ from $P_2$, with frequencies within the filter band and both arriving on detectors A and B. The output currents $i_A$ and $i_B$ from the two sensors will be proportional to the intensity of the light:

$$i_A = K_A [E_1 \sin(\omega_1 t + \phi_1) + E_2 \sin(\omega_2 t + \phi_2)]^2 \tag{8}$$

$$i_B = K_B [E_1 \sin(\omega_1 (t + d_1/c) + \phi_1) + E_2 \sin(\omega_2 (t + d_2/c) + \phi_2)]^2 \tag{9}$$

where $K_A$ and $K_B$ are constants depending on the sensors and $d_1$ and $d_2$ are the optical path differences between the two sensors with respect to the plane wavefront arriving from the star. Developing the square of the fields, it is immediately found that at the output of the sensors there are four components: 1) a d.c. term proportional to the total light flux falling on the detectors; 2) a second harmonic component at frequency $2\omega$; 3) a frequency sum component at $\omega_1+\omega_2$; 4) a frequency difference component, $\omega_1-\omega_2$. Thanks to a low band pass filter installed at the detector output, only the latter passes and reaches a multiplier where the signal coming from the two sensors are multiplied together (i.e. *correlated*). Assuming for simplicity that $\omega_1 \approx \omega_2 = \omega$ (thanks to the narrow band filter in front of the sensors) and considering that $\phi_1$ and $\phi_2$ are independent random variables distributed uniformly, at the end it can be found that these Fourier components contribute to the total signal exiting the multiplier with the following term:

$$c_{12}(d) = K_A K_B E_1^2 E_2^2 \cos[(\omega/c)(d_1 - d_2)] = K_A K_B E_1^2 E_2^2 \cos(2\pi d\theta/\lambda) \tag{10}$$

where $d$ is the separation between the detectors, $\theta$ is the angular separation between the two points $P_1$ and $P_2$ on the star, and $\lambda$ is the mean wavelength of the light passing through the narrow band filter. This relation shows how the correlation of the currents relative to the frequency difference has intrinsically the information about the angular size of the star, even if "distributed" over all the possible angular separations $\theta$ for all the possible couples of points $P_1$ and $P_2$. Integrating this result over all possible pairs of points on the star disc, over all the possible Fourier components within the optical bandpass and over all difference frequencies within the low band pass filter at the output of the sensors [30], it is obtained that the "global" correlation is

$$c(d) = c_o |\gamma_{AB}(\tau)|^2, \tag{11}$$

where $c_o$ depends on the instrumental apparatus and can be determined by a suitable calibration. Following for example [31], it can be found that in the simple case in which the aperture of the telescopes is small with respect to baseline needed to resolve a star having a circular disc of uniform intensity, it is

$$c(d) = \left[\frac{2J_1(\pi\theta_{UD}d/\lambda_o)}{\pi\theta_{UD}d/\lambda_o}\right]^2 \tag{12}$$

where $J_1$ is the first order Bessel function, $\theta_{UD}$ is the angular diameter of the equivalent uniform disc and $\lambda_o$ is the mid band wavelength of the light, and it has been assumed that the light is monochromatic.

This description shows that the information about the angular size of the star can be obtained by the measurement of the correlation between the light intensity fluctuations: in fact, this provides a term which is proportional to the square modulus of the complex degree of coherence of light at the two sensors, and the latter depends on the star angular size. This is what Hanbury Brown and Twiss did, by using the signal provided by two photomultipliers as input to the just described correlator. However, as previously mentioned, there is also the possibility of performing the same measurement using a different approach, based on "counting" the photons arriving on the same sensors.

## 2.2 Measuring coincidences

To find the number of events detected by a photon counting sensor when illuminated by a light beam of intensity $I(r,t)$, we can adopt a probabilistic approach, and say that the average number of events detected in the time interval $[t, t+\Delta t]$ is given by

$$p_1(r,t)\Delta t = \eta\langle I(r,t)\rangle\Delta t \tag{13}$$

where $p(t)$ is a probability density, $\eta$ is the detector quantum efficiency and the intensity is averaged over the ensemble of realization of the field. In case of a stationary field, $\langle I(r,t)\rangle = \langle I(r)\rangle$ and $p(r,t) = p(r) = \eta\langle I(r)\rangle$; so the average number of events detected in a time interval $T$ on detector A or B is given by

$$N_{A/B} = \int_t^{t+T} p_{1,A/B}(r_{A/B},t')\mathrm{d}t' = p_1(r)T = \eta_{A/B}\langle I(r)\rangle T \tag{14}$$

Similarly, we can calculate the number of "simultaneous" (i.e. within the same time interval $\Delta t$) events on two detectors in time intervals $\Delta t$ respectively, obtaining

$$p_2(r_A,t_A;r_B,t_B)\Delta t^2 = \eta_A\eta_B\langle I(r_A,t_A)I(r_B,t_B)\rangle\Delta t^2 \tag{15}$$

where $p_2(r_A,t_A;r_B,t_B)$ is the conditional probability of detecting one event per sensor.

In this case, the average number $N_{AB}$ of coincidences in the same time interval is given by

$$N_{AB} = \int_t^{t+T}\left(\int_{t'-\Delta t/2}^{t'+\Delta t/2} p_2(r_A,t';r_B,t'+\tau)d\tau\right)\mathrm{d}t' = \left(\int_{-\Delta t/2}^{\Delta t/2}\eta_A\eta_B\langle I_A(t)I_B(t+\tau)\rangle d\tau\right)T \tag{16}$$

We can now make use of equation (4) in the case of unpolarized light (that is introducing a ½ factor in front of the modulus of the complex coherence) and of equation (14) for writing

$$N_{AB} = \eta_A\eta_B T\int_{-\Delta t/2}^{\Delta t/2}\langle I_A(t)\rangle\langle I_B(t)\rangle\left[1+\frac{1}{2}|\gamma_{AB}(\tau)|^2\right]d\tau = \frac{N_A N_B}{T}\int_{-\Delta t/2}^{\Delta t/2}\left[1+\frac{1}{2}|\gamma_{AB}(\tau)|^2\right]d\tau \tag{17}$$

It has been shown [32] that in many physical situations the coherence possesses a "cross-spectral purity", which makes the complex coherence reducible to the product of two simpler functions, $\gamma_{AB}(\tau) = \gamma_{AB}(0)\gamma_{AA}(\tau)$. It can be noticed that this simplification transforms the complex coherence in the product of the "spatial" coherence at the two detectors (i.e. measured at null time delay, which is equivalent to have the two telescopes on the same stellar wavefront) with the "temporal" coherence of the light.

By means of equation (17), and moving to count rates $n = N/T$, we get

$$n_{AB} = n_A n_B \left[ \Delta t + \frac{1}{2} |\gamma_{AB}(0)|^2 \int_{-\Delta t/2}^{\Delta t/2} |\gamma_{AA}(\tau)|^2 d\tau \right] \quad (18)$$

Finally, if the sampling time interval is much longer than the coherence time of the light $\tau_o$, $\Delta t \gg \tau_o = 1/\Delta\nu$, this equation simplifies as

$$n_{AB} = n_A n_B \Delta t \left[ 1 + \frac{1}{2} |\gamma_{AB}(0)|^2 \tau_o/\Delta t \right] \quad (19)$$

Inverting equation (19), we obtain

$$\frac{2}{\tau_o} \left( \frac{n_{AB}}{n_A n_B} - \Delta t \right) = |\gamma_{AB}(0)|^2 \quad (20)$$

This last equation is the fundamental result for this experiment, as it describes that it is possible to measure the modulus of the complex coherence by determining the count rates of single events and of coincidences on the detectors. Then, by varying the baseline, it is possible to get more measurements of $|\gamma_{AB}(0)|^2$, which can then be fitted for example with equation (12) to obtain the angular size of the observed object.

It is in some cases more convenient to work with the total number of events and coincidences detected during the observation time *T*. In this case equation (19) can be rewritten, obtaining the so-called $g^{(2)}$ function:

$$g^{(2)}_{AB}(0) = \frac{N_{AB}}{N_A N_B} \frac{T}{\Delta t} = 1 + \frac{1}{2} |\gamma_{AB}(0)|^2 \tau_o/\Delta t \quad (21)$$

This equation is conveniently used to measure $|\gamma_{AB}(0)|^2$, since the total counts is typically the first product of the data analysis.

Equation (19) shows that if there is no coherence between the light at the two detectors, the coincidence rate is $n_A n_B \Delta t$, which is the value expected for uncorrelated streams of photons. However, when there is some coherence, there is an excess of coincidence rate, which depends on the square of the coherence function. This excess of coincidence counts with respect to those expected for the uncorrelated photons is exactly the "signal" that is necessary to measure for making an II experiment. The major issue is to be able to distinguish this contribution from the noise, since the "signal" is typically very small. The standard definition for the S/N in this type of experiment is found considering as main noise contribution the fluctuation in the coincidence rate due to the uncorrelated light: assuming a Poissonian distribution, this noise (standard deviation) equals the square root of the uncorrelated coincidences. Given a total integration time *T*, the number of uncorrelated coincidences is $n_A n_B \Delta t T$, so the S/N ratio is

$$(S/N)_{rms} = \frac{n_A n_B \Delta t \frac{1}{2} |\gamma_{AB}(0)|^2 (\tau_o/\Delta t) T}{\sqrt{n_A n_B \Delta t T}} = |\gamma_{AB}(0)|^2 \frac{\tau_o}{2} \sqrt{\frac{n_A n_B T}{\Delta t}} = |\gamma_{AB}(0)|^2 \frac{\tau_o}{2} \sqrt{\frac{N_A N_B}{T \Delta t}} \quad (22)$$

## 3. EXPERIMENTAL SETUP

To validate the possibility of performing a photon counting stellar II measurement with a baseline of the order of km, we are using the two actually most performing photon counter photometers for astronomical applications, namely Aqueye+ [16] and Iqueye [17]. These two ultrafast photometers are essentially based on the same instrumental scheme: the first was realized for being mounted at the Asiago Copernico telescope, the second was an improved "replica" adapted for being used at the NTT in La Silla. The latter is presently available in Asiago too, so it has been a couple of years ago that we started thinking about the possibility of using both instruments simultaneously on the two available telescopes: one is the 182 cm Copernicus Telescope at Asiago Cima Ekar (T122), to which Aqueye+ is routinely mounted; the other is the 122 cm Galileo Telescope at Asiago Pennar (T122). As already said, these are rather small telescope to actually realize a valuable stellar II experiment; however, the performance of the available instrumentation is such that a reasonable test for validating the technique is actually feasible.

## 3.1 The photon counting photometers

As previously mentioned, Aqueye and Iqueye are based on the same instrumental concept. They both work over a very narrow (few arcsec) field of view, so they can essentially target only unresolved objects in the sky. Light focused by the telescope is sent to a field camera (for having a context view of the pointing area) with the exception of a few arcsec central portion which passes through a small aperture and enters the instrument. Centering the target on this small aperture, allows its light to pass through a focal reducer where a first set of suitable filters can be inserted. Then light is split by a pyramidal mirror in four parts, each corresponding to one quarter of the telescope aperture, and sent through independent optical paths to four photon counting sensors (see Figure 1). Along these four paths additional filters can be inserted, allowing also to realize simultaneous spectral photometric analysis. The used sensors are the extremely performing single photon avalanche diodes (SPADs) detectors provided by MicroPhoton Devices (MPD, Italy): they give a 30-50 ps time resolution on the photon detection, have a 50-100 dark count/s, allow a 10-12 MHz maximum count rate (even if the linear regime is of the order of 4-5 MHz), have an approximately 80 ns dead time after a detection, and provide a peak quantum efficiency of about 60% [33][34].

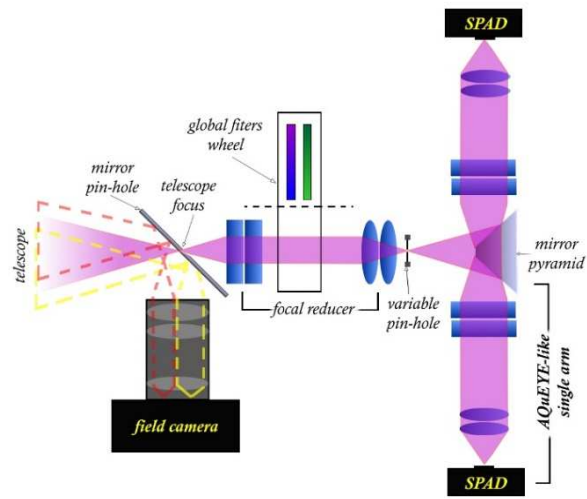

Figure 1. Schematic of the optical layout of Aqueye (and Iqueye).

The pulses provided by each SPAD at the photon detections are sent to a Time do Digital Converter (TDC) board made by CAEN (Costruzioni Apparecchiature Elettroniche Nucleari, Italy), and the 21-bit digital information on the arrival time (24.4 ps time bin) is sent via fiber optics to the acquisition server. The server bus and data storage are the actual bottleneck of the instrument, limiting the counting rate in a linear regime to at most 2 MHz. The TDC provides a timing information modulus of ~51.2 µs, so it is necessary to provide an external accurate clock to get the complete time information over the hours of an observation run. For this, the instruments make use of an external rubidium clock (Stanford Research Systems, USA) and a GPS unit (Trimble, USA), which sends pulses at a well defined frequency to the same TDC. An ad hoc software then uses the information provided by these two external references for obtaining an extremely stable and accurate absolute reference clock on which the detected photon time events are then tagged. All time tags are finally stored on an external archive, where they can be retrieved for any possible data reduction. The actual limit in the time tag resolution is due to the system electronics, and we estimated it to be of the order of 100 ps: this is the "relative" error on the pulse time, but not the "absolute" one. The latter is of the order of 0.5 ns (with respect to UTC), and is related to the intrinsic time error of the GPS pulse and the statistics associated to the number of pulses detected from the GPS (one per second) during the observation time. A schematic of the instrument electronics system is shown in Figure 2.

The characteristics of these two instruments (see Figure 3) make them perfectly suitable for performing nano- and sub-nanosecond time resolution, as already demonstrated by the best ever results obtained in the timing analysis of the optical pulsars, the fastest variable objects in the sky. Moreover, the data storage allows an incredible versatility in post-processing data analysis, allowing to avoid a "physical" connection between the two telescope and to define for example any time bin of interest from seconds down to nanoseconds. This is a fundamental characteristic in view of an II measurement over a km-long baseline, and a great advantage with respect to the more classical techniques of real-time cross correlation (used for example by HBT), in which the data collected from the telescopes have to be electronically sent to a correlator, and only the final product of the correlation is provided, losing all the original data and not allowing any sort of possible re-elaboration or correction of the analysis.

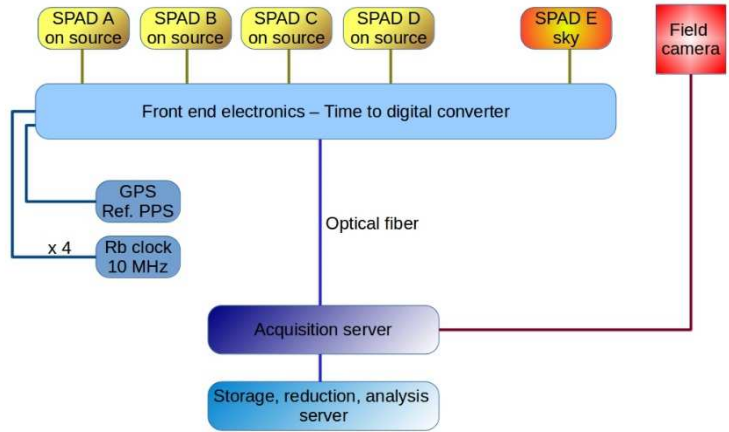

Figure 2. Schematic view of the Aqueye+ and Iqueye acquisition and timing system.

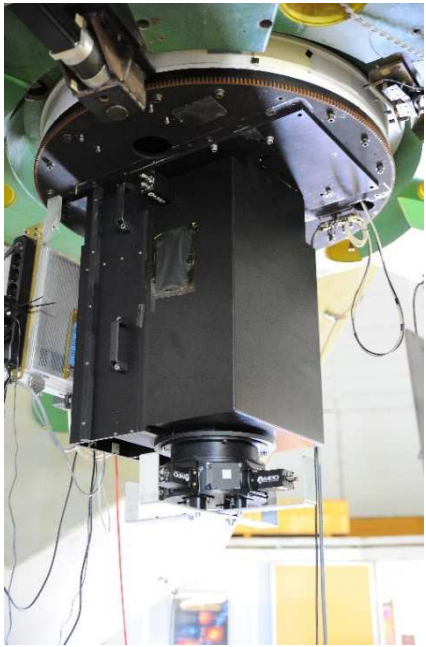
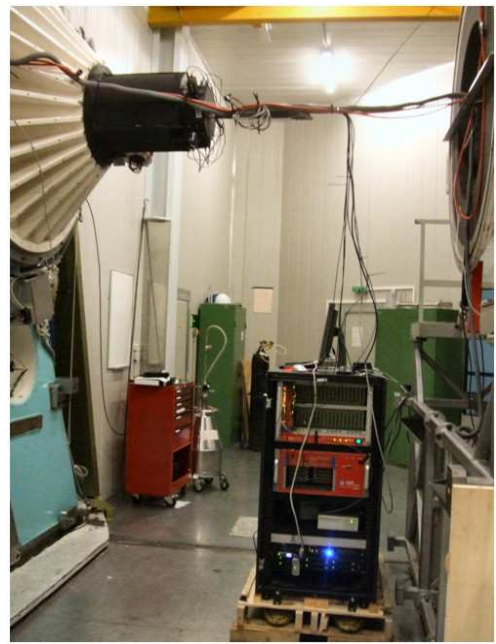

Figure 3. Left: Aqueye+ mounted at the T182 Copernicus telescope in Asiago. The MPD SPAD detectors are visible at the bottom. Right: Iqueye mounted at NTT. The rack at the base hosts the whole electronics system.

Given some mechanical interference problems in the direct mounting of Iqueye at the T122, for this experiment we adopted a different mounting philosophy for Iqueye. By means of a suitable optical bench connected to the telescope rear flange, a fiber is mounted in the telescope focal plane; the target light, focused on the fiber end, is so brought to the other fiber end which is connected to the entrance input of the instruments. In this way, not only the instrument installation is made much simpler, but it is also possible to maintain the instruments in better environmental conditions, mainly of temperature and humidity, so largely reducing potential problems related to variations of the ambient conditions (for example, SPADs can stop working at ambient temperatures below –15°C, a situation not uncommon in winter time). Aqueye is presently mounted through its flange at the T182 rear plate (see Figure 3 left), but also for this second instrument it is foreseen a similar soft mounting in the near future. The main constraints in this case are the need to work remotely and the limitation in manpower resources. Clearly, the fiber link is less efficient than a direct mounting of the instruments to the telescope focus, and we estimated a loss of about 15-20% in transmission. A detailed description of this new installation configuration can be found in [24].

### 3.2 Implementation of the Asiago T122-T182 Intensity Interferometry experiment

The two telescopes foreseen for this experiment, the T122 and T182, are the largest at the Asiago Observatory (Italy); they are separated by about 3.9 km, with a significant East-West component (see Figure 4). The geographic and cartesian geocentric coordinates of the two telescopes, measured with a GPS receiver and referred to the intersections of their hour angle and declination axes by means of laser-assisted metrology, are reported in Table 1. With such a baseline, the expected angular resolving capability of this experiment is of the order of tens of µas (see Figure 5).

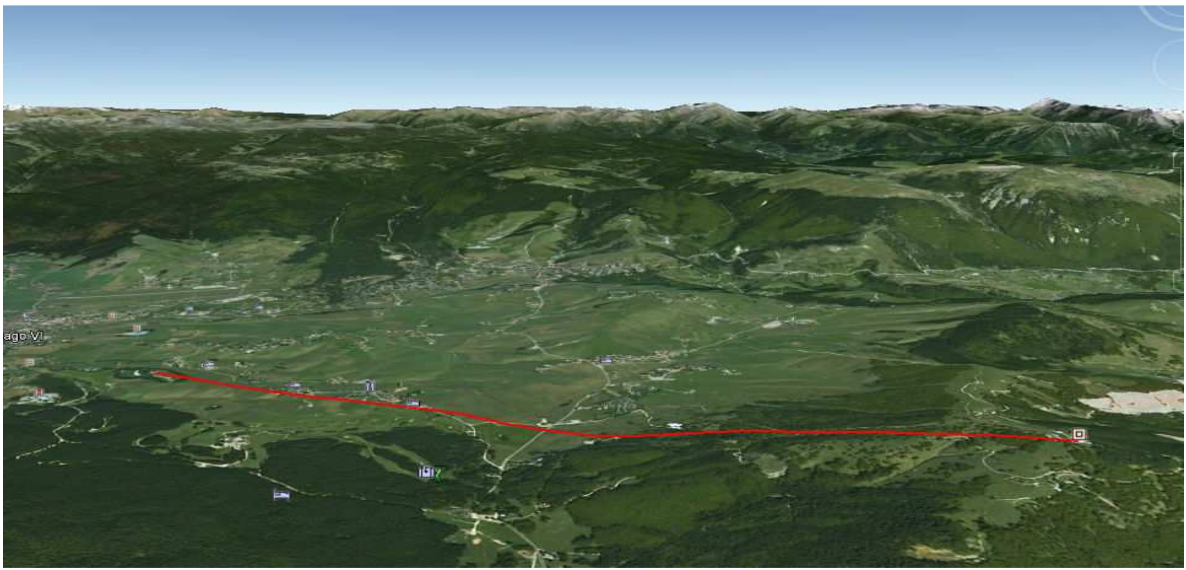

Figure 4. Picture (from Google Earth) of the Asiago area. The T122 and T182 telescopes are at the extremes of the red line: on the left, the Galileo telescope T122 at the Pennar station; on the right, the Copernicus telescope T182 at Cima Ekar.

It has to be reminded that, being the two telescopes at fixed locations, generally they will not be positioned on the same wavefront of the incoming stellar light, where photons are correlated. Since equations (19)-(22) can be used under the $\tau = 0$ condition, it is necessary to artificially compensate this effect by introducing a delay time $T_d$ at the time of arrival of the photons collected at one of the two telescopes; clearly, this delay depends on the relative orientation between the baseline and the propagation direction of the light, and changes in time during an observation. This compensation can be done "easily" with our setup, since it can be added by software during the data reduction process. The light travel time delay between the two telescopes at the beginning of an observation can be very accurately determined by using existing software. For example, it is possible to take the UTC times recorded for the first detected events and to "bring" them to the barycenter of the solar system with TEMPO2 software [35][36]. By comparison of the times to be added for this barycentrization on the two observing sites, and considering also the additional delay introduced by electronics cables and the Iqueye fiber link, the initial time delay is determined. Then, thanks to the proximity of the two observing sites, the delay variation during the observation time can be determined by taking in account the variation of the projected baseline with respect to the direction of the star.

In addition, the fixed telescope baseline associated with the Earth rotation provides a variability in the projected baseline length. In fact, for a source at small elevation in the East/West direction, the projected baseline is approximately half the full baseline, while for a star at zenith in practice it coincides with the actual baseline. This extreme cases demonstrate that the projected baseline can vary by a factor ~2 and in principle enables us to sample the correlation of the signal as the baseline changes during an observing night.

Table 1. Coordinates and baseline of the Copernicus T182 and Galileo T122 telescopes in Asiago. Coordinates refer to the intersections of the hour angle and declination axes.

| GEOGRAPHIC AND CARTESIAN GEOCENTRIC COORDINATES OF THE COPERNICUS (T182) AND GALILEO (T122) TELESCOPES IN ASIAGO | | | |
|---|---|---|---|
| T182 geographic | | T122 geographic | |
| Longitude | 11 34 08.81 E | Longitude | 11 31 35.14 E |
| Latitude | 45 50 54.47 N | Latitude | 45 51 59.22 N |
| Elevation | 1410 m | Elevation | 1094.6 m |
| T182 cartesian | | T122 cartesian | |
| X | 4360966.0 m | X | 4360008.6 m |
| Y | 892728.1 m | Y | 889148.3 m |
| Z | 4554543.1 m | Z | 4555709.2 m |
| **BASELINE BETWEEN THE TWO TELESCOPES** | | | |
| Baseline (T182-T122) | | | |
| DX | 957.4 m | | |
| DY | 3579.8 m | | |
| DZ | -1166.1 m | | |
| B=(DX$^2$+DY$^2$+DZ$^2$)$^{1/2}$ | 3884.8 m | | |

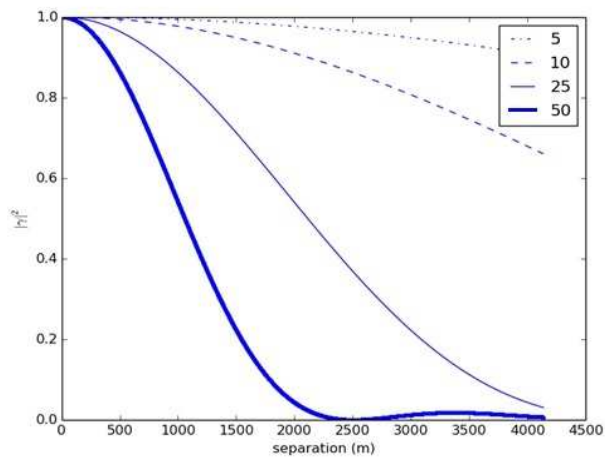

Figure 5. Square modulus of the mutual degree of coherence $|\gamma_{AB}(0)|^2$ for visible light (500 nm) as a function of the separation between the telescopes. The different curves are labeled with the apparent angular size of the star in microarcseconds.

## 4. EXPERIMENT DESCRIPTION

To understand what can actually be done with the just described experimental setup, we need to have in mind that our instruments make a "parallelization" of the signal after the telescope focusing, sending light independently on four separate SPADs; moreover, we store the time tags of all detected events with a *relative* rms time accuracy (that is relative to the "internal" clock) of the order of 100-200 ps, and an *absolute* rms time accuracy (that is relative to UTC) of the order of 0.5 ns. Thanks to these characteristics, we can use Aqueye+ at T182 to measure the $g^{(2)}_{d=0}(0)$, that is the $g^{(2)}$ function at zero baseline, getting the maximum correlation ($|\gamma_{d=0}(0)|^2 = 1$). In fact, we can sum up the counts from only two SPADs, and correlate them with the counts from the other two SPADs. In this way, we are correlating photons on the same wavefront from two telescope apertures separated by about 1 m, which gives an essentially null baseline. Moreover, during this measurement, we can use the relative rms time accuracy of the instrument, let's say 200 ps, as resolving time, since we are using the same accurate clock for the two telescope aperture photon streams. Then we can simultaneously use Aqueye+ at T182 and Iqueye at T122 to measure $g^{(2)}_d(0)$, that is the $g^{(2)}$ function with a baseline $d$ corresponding to the telescope separation projected on the wavefront, obtaining the correlation $|\gamma_d(0)|^2$. In this case the minimum resolving time will actually be the absolute rms time accuracy, that is 0.5 ns[†], since the two clocks have to be synchronized to the absolute time provided by GPS. Moreover, in this case the projected baseline will not be constant, depending on the star position with respect to the telescopes, and roughly varying in the range [2-3.9] km. But looking at stars with angular size of the order of ~mas, with such a long baseline there will be practically no correlation ($|\gamma_{d \sim km}(\tau)|^2 = 0$, see Figure 5). In practice, what we can do with our present setup is measuring two points of the correlation curve, the maximum and the minimum. As better described in the following, it is extremely difficult with our experiment to take other intermediate measurements, so being able to actually sample the $|\gamma_{AB}(0)|^2$ and to determine the stellar angular size. However, if we can demonstrate this capability with such a limited instrumentation, it is clear that the technique could be successfully applied to larger telescope and more updated photon counting photometers to get an actual photon counting stellar II.

In order to have an idea about the values of our measurement, we can make use of the previously listed formulas. Let's start considering the case of maximum correlation, as in the case of zero baseline ($|\gamma_{d=0}(0)|^2 = 1$). Presently, the narrowest filter we can use is the O[III], centered at $\lambda = 501$ nm with a spectral bandwidth of $\Delta\lambda = 1.3$ nm; this corresponds to $\Delta\nu = 1.55$ THz and to a coherence time $\tau_o = 0.644$ ps. With these inputs and using a time bin of $\Delta t = 0.2$ ns, from equation (21) we obtain that the expected value of the $g^{(2)}$ function is $g^{(2)}_{d=0}(0) = 1.00161$. Since the smallest value for $g^{(2)}$ is $g^{(2)}_{AB}(0) = 1$ in case of uncorrelated light, we see that just to discriminate these two extreme cases we need to perform a measurement with an accuracy better than $1.6 \cdot 10^{-3}$ (obviously, this accuracy has to be significantly improved if sampling the $g^{(2)}$ function is requested to determine the actual star angular diameter).

We can realistically assume a count rate of the order of ~2 MHz (i.e. ~1 MHz on two SPADs) at T182, and ~1 MHz at T122, which can be obtained by pointing at bright stars (e.g. Deneb, the brightest star of the Cygnus constellation) with the O[III] filter inserted. Considering a total integration time of half an hour, from equation (21) in case of null baseline it is found that the expected number of coincidences is about $3.61 \cdot 10^5$. Of all these coincidences, only 0.161%, that is 579, are due to the light correlation, all the others being random coincidences associated to the uncorrelated portion of the light. On the basis of equation (22), it can be found that the (S/N)$_{rms}$ under these conditions is 0.97, so actually providing a measurement at the limit of meaningfulness. However, we can easily increase the integration time to rise the (S/N)$_{rms}$ because, given the experimental condition previously described, there is no need to perform a single "continuous" observation to obtain the final result: we can simply calculate the total counts per sensor and the total coincidences detected during several different observing runs, sum them, and finally calculate $g^{(2)}$ and (S/N)$_{rms}$. This is due to the fact that we operate either with a null baseline (full correlation), or with a very long baseline (no correlation), and thanks to our data storage capability we can make all the data analysis in post-processing. Table 2 provides an indication of the expected results at exposure times longer than half an hour. In addition, since our measurable is the $g^{(2)}$ function, we can also estimate the standard deviation associated to this measurement by simple propagation assuming Poisson distributions of the variables, and negligible error on integration time and time bin definition. The rms error on $g^{(2)}$ is simply

---

[†] Actually, since the photons collected at the T122 are injected in a 10 m fiber before reaching Iqueye, there is a broadening of the pulse due to the fiber modal dispersion. Considering the fiber characteristics, the total broadening is about ±0.22 ns, which brings the rms time accuracy to about 0.7 ns.

$$\sigma_g = g^{(2)}_{d=0}(0) \sqrt{\frac{1}{N_{AB}} + \frac{1}{N_A} + \frac{1}{N_B}} \approx \frac{1}{\sqrt{N_{AB}}} \qquad (23)$$

which, with the given coincidence count rate, is $\sigma_g = 1.18 \cdot 10^{-3}$ for one hour of observation. Given these numbers, we see that if no other systematic error is present, it is possible to make a measurement which allows to discriminate the totally correlated case with respect to the totally uncorrelated one at $3\sigma_g$ with a total observation time of the order of 5 hours or more.

Table 3 shows what are the $2\sigma$ expectation when sampling $|\gamma_{AB}(0)|^2$ assuming still 1 MHz count rate on each sensor, but using an ultra-narrow 1 nm band H-$\alpha$ filter ($\lambda = 656$ nm, coherence time $\tau_o = 1.43$ ps) with a time bin $\Delta t = 0.5$ ns. This corresponds to a possible measurement to do when varying the baselines for making an actual measurement of the stellar angular size. Assuming larger telescopes (for example, with an 8 m telescope the count rate can be about 20 times larger in proportion to the T182), this would allow to measure stars of apparent visible magnitude up to 4. It is clear from these values that a sampling of $|\gamma_{AB}(0)|^2$ could be done under these conditions.

Table 2. Estimates of the expected "random" and "correlation" simultaneous detections, and of the consequent signal-to-noise ratio and statistical error on $g^{(2)}_{d=0}(0)$, for the experiment setup described in the text. The three cases, 0.6, 3 and 5 hours, correspond respectively to the cases in which the statistical errors on $g^{(2)}_{d=0}(0), \sigma_g, 2\sigma_g, 3\sigma_g$ are smaller than the difference between the totally correlated and totally uncorrelated $g^{(2)}$.

| Exposure time (hours) | 0.6 | 3 | 5 |
|---|---|---|---|
| Expected "random" simultaneous detections | 4.32E+05 | 2.16E+06 | 3.60E+06 |
| Expected "correlation" simultaneous detections | 695 | 3475 | 5792 |
| (S/N)$_{rms}$ | 1.06 | 2.36 | 3.05 |
| $\sigma_g$ | 1.52E–03 | 6.81E–04 | 5.28E–04 |

Table 3. Estimated signal-to-noise ratio and statistical error on $g^{(2)}_{AB}(0)$ function, in the case of an ultra-narrow band H-$\alpha$ filter ($\lambda = 656$ nm, $\Delta\lambda = 1$ nm) and resolving time $\Delta t = 0.5$ ns, assuming that the difference between the expected $g^{(2)}_{AB}(0)$ and the uncorrelated one is larger than $2\sigma_g$.

| Assumed value of $|\gamma_{AB}(0)|^2$ | 0.8 | 0.6 | 0.4 | 0.2 |
|---|---|---|---|---|
| Expected $g^{(2)}_{AB}(0)$ | 1.001148 | 1.000861 | 1.000574 | 1.000287 |
| (S/N)$_{rms}$ | 2.18 | 2.16 | 2.04 | 2.11 |
| $\sigma_g$ | 5.28E–04 | 3.99E–04 | 2.82E–04 | 1.36E–04 |
| Exposure time (hours) | 2 | 3.5 | 7 | 30 |

## 5. EXPERIMENT STATUS

We performed the first experimental/commissioning run of the Asiago Intensity Interferometer in July 2015, and another preliminary run in January 2016 (other two observing runs were foreseen in March and in July 2016, but bad weather conditions did not allow to make successful measurements). We observed a few stars of early spectral type (O, B and A), one late spectral type star and Deneb. Deneb and HR 5086 were targeted for trying a preliminary measurement of $g^{(2)}_{d=0}(0)$ with Aqueye+ at T182. In this case, we wanted to see possible differences in the correlation between the signal detected by two SPADs observing stars with different colors and magnitudes. The other targets are blue main sequence stars that, given their visible magnitude (hence distance) and expected radius (2-8 solar radii), should have an angular diameter of the order of ~10 µas and then be potentially resolvable on a few km baseline (see Figure 5). A log of the two observing runs is reported in Table 4; Figure 6 shows the light curves of Deneb simultaneously acquired on July 31st with Aqueye+ and Iqueye. We report here only about the very preliminary analysis of the data acquired on Deneb on July 2015 at T182, since further work is necessary in order to get more consolidated results.

Table 4. Summary information of the first two experimental/commissioning runs of the II experiment.

**LOG OF THE FIRST EXPERIMENTAL OBSERVING RUNS OF THE ASIAGO INTENSITY INTEFEROMETER**

| July 2015 | | | January 2016 | | | |
|---|---|---|---|---|---|---|
| Iqueye@T122 with optical fiber | | | Iqueye@T122 with optical fiber | | | |
| Aqueye+@T182 | | | Aqueye+@T182 | | | |
| Targets | Spec. Type | V mag | Date | UTC | Filt. | Duration (s) |
| Deneb | A2Ia | 1.25 | Jul 31, 2015 | 21:28:46 | V | 900 |
| Deneb | | | Jul 31, 2015 | 21:50:45 | V | 1800 |
| Deneb | | | Jul 31, 2015 | 22:28:09 | V | 900 |
| Deneb | | | Jul 31, 2015 | 23:00:21 | H-$\alpha$ | 1800 |
| BD+62 249 | O9.5V | 10.2 | Aug 1, 2015 | 01:23:49 | V | 900 |
| BD+62 249 | | | Aug 1, 2015 | 01:46:50 | V | 900 |
| BD+62 249 | | | Jan 16, 2016 | 21:52:41 | | 3600 |
| BD+60 552 | B9V | 10.9 | Jan 17, 2016 | 21:02:55 | | 3600 |
| BD+58 629 | A0/2V | 10.7 | Jan 18, 2016 | 01:48:40 | | 1800 |
| BD+58 629 | | | Jan 18, 2016 | 02:23:21 | | 1800 |
| HR 5086 | K5V | 6.2 | Jan 18, 2016 | 04:02:15 | V | 3600 |

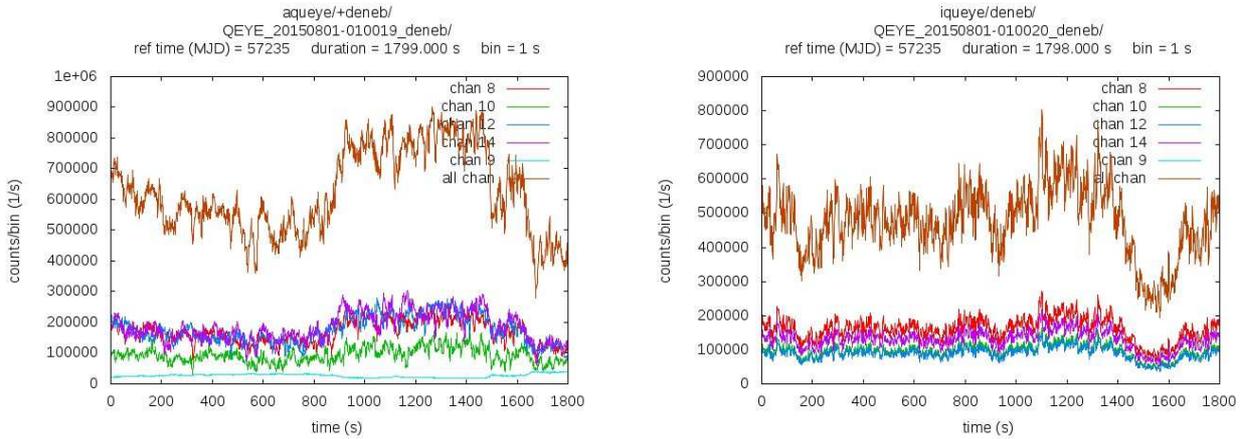

Figure 6. Light curves of the observations of Deneb acquired with the H-$\alpha$ filter on July 31, 2015, starting at 23:00:21 UTC (the Iqueye observation started 1 second later). *Left*: Aqueye+. *Right*: Iqueye. The bin time is 1 s. Count rates of each SPAD (channels 8 through 14) and of the sum on all on-source channels are shown. Channel 9 is the SPAD detector monitoring the sky (switched off in Iqueye). The acquisition with Aqueye+ was obtained inserting a neutral attenuator filter to limit the count rate. The broad oscillations visible in both images are caused by the not perfect sky conditions (passage of clouds and veils).

To acquire confidence with the data analysis and with the ad hoc written correlator software[‡], we decided to measure first the correlation of the photons collected with the observation of Deneb at T182 with Aqueye+ only: in practice we wanted to determine the $g^{(2)}_{d=0}(0)$ value to verify that the correlation in these condition is the maximum ($|\gamma_{d=0}(0)|^2 = 1$). This choice was due to the fact that this is the simplest measurement that can be done, since there is no need of correlating the signals from the two telescopes, there is just one reference internal clock, and so there are less problems to solve and criticalities in the data analysis. In particular, we used the total counts and coincidences measured by each single SPAD of Aqueye+ as inputs for equation (21): in this way, we determined the values of $g_{d=0}^{(2)}(0)$ for all the possible combinations of the four SPADs and then we could calculate $|\gamma_{d=0}(0)|^2$. Actually, the telescope sub-apertures from which each SPAD collects photons are separated by less than one meter, and with the Deneb apparent angular diameter of about 2 mas, the signals from the SPADs are expected to be significantly correlated. For this observation, an H-$\alpha$ filter was used ($\lambda$ = 656 nm, $\Delta\lambda$ = 3 nm), and we set $\Delta t$ = 0.2 ns; clearly the mirror sub-apertures are on the same wavefront, so that $\tau$ = 0.

---

[‡] A brief description of the correlator software is provided in the following subsection.

With these parameters and assuming maximum correlation, $|\gamma_{d=0}(0)|^2 = 1$, from equation (21) we expect to find $g_{d=0}^{(2)}(0) = 1.0012$. The results we obtain are summarized in Table 5. It can be seen that the reported values, that are averages of the corresponding values obtained dividing the total observation in 20 segments, are in line with expectations only for the couple of SPADs B-C, although the statistical error is anyway too large to draw any significant conclusion: have in mind that with these count rates and exposure time, the expected signal to noise ratio is (S/N)$_{rms}$ = 0.14, showing that either longer acquisition and/or higher count rates are necessary for being confident about the obtained result. In all the other cases, however, the obtained values have no physical meaning, also considering a $3\sigma_g$ error. In fact, $g_{d=0}^{(2)}(0)$ has always to be larger than one, while here we are about 6-12% smaller, and the standard deviation is of the order of 1-1.5%. We have investigated several possible causes of this result: detector dead time, possible electronics delay, software errors, short resolving time, but we have not found any satisfactory explanation yet. It should be mentioned that a similar test was performed in 2010 with Iqueye at NTT [23], observing $\zeta$ Orionis without narrow band filter along the optical path: the instrument and the experimental conditions were different, as well the used correlator software, but also in that case we obtained a $g_{d=0}^{(2)}(0)$ value about 10% smaller than expected. Thus, it is not evident if there is some systematic issue with our instrumentation, or if there are other possible explanations for justifying the loss of about 10% of coincidences. What we also wanted to do was to repeat the observations possibly for longer times and higher count rates to get some better statistics: unfortunately, it is not so simple to get available time at the two telescopes simultaneously, so the actual possibilities to perform this measurement are just two or three per year, and even after that we cannot control the weather conditions.

Table 5. Measured $g_{d=0}^{(2)}(0)$ values obtained from the signals by the four SPADs of Aqueye+ at T182 during an observation of Deneb with the H-$\alpha$ filter taken on July 31, 2015 at 23:00:21 UTC. The adopted bin time is $\Delta t$ = 200 ps. The reported values are the mean and standard deviation of the mean of the measurements performed having divided the observation in 20 segments. The approximate baseline is calculated taking as reference points the corners of the square inscribed in a circle with half diameter of the telescope primary mirror.

| SPADs | Baseline (m) | $g_{d=0}^{(2)}(0) \pm \sigma_g$ |
|---|---|---|
| A-B | 0.6 | 0.908 ± 0.016 |
| A-C | 0.9 | 0.901 ± 0.009 |
| A-D | 0.6 | 0.937 ± 0.010 |
| B-C | 0.6 | 1.024 ± 0.016 |
| B-D | 0.9 | 0.879 ± 0.017 |
| C-D | 0.6 | 0.926 ± 0.009 |

## 5.1 Work in progress

At present we are also tackling the cross-correlation of the signals obtained with the two telescopes. Things are more complicated with respect to the zero baseline case not only by the need of having a common clock on the two instruments, but also by the issue of compensating for the baseline variation during the observing time.

The software correlator (written in Linux bash shell and Fortran) takes the datasets stored in the Aqueye+ and Iqueye archives, which are a list of UTC times corresponding to the event detections, and first generates ~90-second long light curves of the signal: these are arrays of the detected event times, binned at the resolving time $\Delta t$. Then it stores only the sequential numbers of the non-zero time bins, so largely reducing the size of the data array (to give an idea about the data volume, a 10-minute acquisition archived dataset, corresponding to several billion photons at these count rates, occupies more than 10 GB of disk space). Before correlating these arrays, a variable time delay $T_d(t)$ has to be added to one of the two light curves, to bring the two telescopes on the same star wavefront. For this, the first data acquired on each telescope are referred to the solar system barycenter with suitable software, and the difference in time between the two barycentrizations provides the initial time delay $T_d(0)$. Considering that the wavefront at the telescopes on average rotates by approximately $7 \cdot 10^{-5}$ radians every second and that the light travel time between the two telescopes is ~10 µs, to maintain the correlation within a time bin, for example, $\Delta t = 1$ ns, $T_d(t)$ has to be varied by ~1 ns every 1.5 seconds. At this point, after the delay compensation, it is possible to look for coincidences in the series of sequential numbers from the two instruments. This Aqueye+/Iqueye light curve software correlator is conceptually similar to the Iqueye software correlator described in [23], with the difference that here the correlation is performed on the binned light curves and not directly on the photon arrival times.

The computational time needed to reduced and analyze the whole dataset is significant. Despite the efficiency of the algorithm, processing a 10 minute dataset requires a few tens of minutes on a standard workstation. Several hundreds of GB of data have been already acquired in the two experimental observing runs of July 2015 and January 2016, and need to be analyzed. Eventually, all the preparatory work and tests done up to now and presented in this paper are of crucial importance to understand how to control the whole system and to have the intensity interferometer properly working.

## 6. CONCLUSIONS

We have started an intensity interferometer experiment to test the capability of measuring the star diameters by exploiting for the first time the particle nature of light. This experiment is just a validation test, since the telescope apertures are too small to allow to sample the normalized correlation function $|\gamma_{AB}(0)|^2$ that would allow to determine the star diameter. The available photon counting photometers are the presently best available, allowing to time tag photons with a sub nanosecond accuracy, and to store all the data for making a post-process data analysis. Thanks to these characteristics, it is possible to use two telescopes which are separated by 3.9 km without the need of any physical connection between the two instruments.

Presently, only very preliminary data are available, relative to the measurement of the zero baseline $g^{(2)}_{d=0}(0)$ function. The measurement is only partially in line with the expectations, showing that there are still some experimental issues to be solved before confirming the goodness of the measurement. However, the theory confirms that with the performance of the presently available instrumentation, reasonable results can be obtained, and we are working on getting them.

## ACKNOWLEDGEMENTS

We would like to thank all the staff of the Asiago Cima Ekar and Pennar observing stations for their continuous support and effective collaboration, essential for the success of this project. We thank also Giovanni Ceribella, Alessia Spolon and Arianna Miraval Zanon for their help during the two observing runs. This work is based in part on observations collected at the Copernicus telescope (Asiago, Italy) of the INAF-Osservatorio Astronomico di Padova under the program "Timing ottico della Crab pulsar ad elevatissima risoluzione temporale con Aqueye+". This research has been partly supported by the University of Padova under the Quantum Future Strategic Project, by the Italian Ministry of University MIUR through the programme PRIN 2006, by the Project of Excellence 2006 Fondazione CARIPARO, and by the INAF-Astronomical Observatory of Padova under the two grants "Osservazioni con strumentazione astronomica ad elevata risoluzione temporale e modellizzazione di emissione ottica variabile" (2014) and "Installazione di Aqueye+ al telescopio Copernico" (2015).